\documentclass[twocolumn,showpacs,preprintnumbers,amsmath,amssymb,floatfix]{revtex4}
\usepackage{graphicx} 
\usepackage{color} 
\usepackage{dcolumn} 
\usepackage{amsmath}
\usepackage{amssymb} 
\usepackage{latexsym} 
\usepackage{bm}
\newcommand{\be}{\begin{equation}} 
\newcommand{\bea}{\begin{eqnarray}}
\newcommand{\bdm}{\begin{displaymath}} 
\newcommand{\edm}{\end{displaymath}}
\newcommand{\eea}{\end{eqnarray}} 
\newcommand{\ee}{\end{equation}}
 
\newcommand{\ra}{\rangle}
\newcommand{\Neel}{N\'{e}el } 
\newcommand{\im}{{\rm i}}

\begin{document}
\title{Influence of an inter-chain coupling on  spiral ground-state correlations in
frustrated 
spin-1/2 $J_1$-$J_2$ Heisenberg chains}
\author{Ronald Zinke$^a$,  Stefan-Ludwig Drechsler$^b$,
and Johannes Richter$^a$ }
\affiliation{$^a$Institut f\"ur theoretische Physik, Universit\"at Magdeburg, 
PO Box 4120, Germany
\\ $^b$Leibnitz-Institut f\"ur Festk\"orper- und Werkstoffforschung (IWF) D-01171 Dresden, P.O. Box 270116, 
Germany}
\date{\today}

\begin{abstract}
We investigate the influence of an 
inter-chain coupling on the spiral ground state
correlations of a 
spin-1/2 Heisenberg
model consisting of a two-dimensional  array of coupled chains with nearest
($J_1$)
and frustrating next-nearest neighbor ($J_2$) in-chain exchange couplings.  
Using the coupled-cluster method we calculate the transition point between
the commensurate and the incommensurate (spiral) ground state as
well as the quantum pitch angle of the spiral ground state. 
In addition, we provide
a simple empirical formula which describes the relation between the quantum pitch
angle and the frustration parameter $J_2/J_1$.       
\end{abstract}

\pacs{75.10.Jm, 75.10.Pq, 75.40.Mg, 75.50.Ee}

\maketitle

\section{\label{intro}Introduction}
In recent years strongly frustrated quantum magnets exhibiting exotic 
ground state
(GS) phases
have been intensively investigated 
both theoretically and experimentally, 
see e.g. Refs.~\onlinecite{science,diep04,lnp04}. 
At zero temperature all transitions between GS phases are driven purely by the 
interplay 
of quantum fluctuations and 
the competition bet\-ween interactions (e.g., frustration), see e.g.\ 
Refs.~\onlinecite{sachdev99,sachdev04,misg04,Mikeska04,Richter04}. 
Particular attention has been paid to one-dimensional 
(1D)
$J_1$-$J_2$ quantum Heisenberg
models,
which have been studied theoretically with much success
over the last two decades, see Ref.~\onlinecite{Mikeska04} and references therein.
From the experimental side 
recent investigations  have shown that  edge-shared chain
cuprates build a special family of frustrated quantum magnets which  can be
described by a quasi-1D $J_1$-$J_2$ Heisenberg model.
Among others we mention here   LiVCuO$_4$, LiCu$_2$O$_2$,
NaCu$_2$O$_2$, Li$_2$ZrCuO$_4$, and Li$_2$CuO$_2$
\cite{gibson,matsuda,gippius,ender,drechs1,
drechs3,drechs4,park,drechsQneu,malek,tarui},
which were identified   as quasi-1D frustrated 
spin-{1/2} magnets 
with ferromagnetic nearest-neighbor (NN) in-chain $J_1$ and antiferromagnetic 
next-nearest-neighbor (NNN)
in-chain interactions
$J_2$.
These cuprates have attracted much attention due to strong quantum effects and
the observation of incommensurate spiral (helical) spin-spin correlations at 
low temperatures.
Among these materials  Li$_2$ZrCuO$_4$ 
and Li$_2$CuO$_2$ are of particular interest, since 
these
compounds are
found to be near a quantum critical point \cite{drechsQneu,malek}.

The 1D frustrated 
spin-{1/2} $J_1$-$J_2$ Heisenberg model may serve 
as the simplest model to describe some important features of 
such
materials.
The GS properties of the model in the
classical limit are well known, 
i.e.\ when the 
spin quantum number  $s \to \infty$. In this case
the GS does exhibit  a second-order
transition from a  collinear phase (ferro or antiferro) to a non-collinear phase
with spiral correlations along the chains 
at $J_2=|J_1|/4$. 
For $J_2\ge |J_1|/4$ the  classical spiral (pitch) angle $\alpha_{\rm cl}$ is 
given by 
\be \label{a_cl}
\alpha_{\rm cl}=\arccos\left( -J_1/4J_2\right).
\ee
Similar expressions can be derived in the presence of further
couplings $J_3$, $J_4$, etc., however, this more general case 
will not be considered 
here for the sake of simplicity.
Note that in the classical limit neither the pitch angle $\alpha_{\rm cl}$ 
nor the transition point 
$J_2=|J_1|/4$ depend on the inter-chain coupling $J_{\perp}$.
Such a classical relation 
as Eq.~(\ref{a_cl}) between $J_1$, $J_2$, $\ldots$ 
and the    pitch angle $\alpha_{\rm cl}$
has been used to justify 1D
parameter sets obtained from fitting $\chi(T)$
data 
\cite{capogna2005} or by LDA mapping procedures \cite{mazurenko2007}. 
However, ignoring this way
sizable quantum effects, such an 
approach
is not very convincing.
Furthermore, in real materials additional terms in the Hamiltonian such as
anisotropy or exchange coupling between the chains might be of
relevance to yield a quantitative theoretical description of the  experimental
results. The existence of helical long-range order at low temperatures
makes the importance of the inter-chain coupling evident.
However, from the theoretical side such extended models so far are much less
studied than the 
pure 1D-''parent'' models.

Therefore, in the present paper we focus on the discussion of the effect of
the inter-chain coupling
$J_\perp$ on
the GS  spin-spin correlations  in frustrated
spin-1/2 Heisenberg
model consisting of a
two-dimensional (2D) array of frustrated chains coupled by $J_\perp$.
In particular, we 
will
discuss GS's with incommensurate
spiral (i.e. noncollinear) correlations. 
Though, meanwhile many papers exist dealing with  spiral correlations in the
strictly 1D system
($J_\perp=0$)
\cite{Hamada,bursill,krivnov96,white96,aligia00,dima06,dima07}, the influence of the
inter-chain coupling on the pitch angle to 
the best of our knowledge has not been discussed so far.

The theoretical treatment of frustrated quantum antiferromagnets is far 
from being trivial. Although one can find exact GS's  in 
some exceptional cases, see e.g.\ Refs.~\onlinecite{Shastry,maju,Hamada,dim_pla,prl02}, 
standard many-body methods may fail 
or become computational infeasible. For instance, the quantum Monte Carlo techniques 
suffer 
from the minus-sign problem
in frustrated systems.
The density-matrix 
renormalization group (DMRG)
successfully used to discuss
spiral correlations in 1D magnets \cite{bursill,white96} is essentially  
restricted to 
1D systems, at least in  the 
present state of the art. 
Also the exact diagonalization technique used in Ref.~\onlinecite{aligia00} to find
the pitch angle for the
1D problem would be 
limited to extremely small chain lengths when a finite
inter-chain coupling should be considered.
A method  fulfilling the requirement  to be able to deal with frustrated spin 
systems 
at  any 
dimension, 
including magnetic systems with incommensurate spiral
GS's, is 
the coupled cluster method (CCM). 
This method was already used for the strictly 1D
$J_1$-$J_2$ Heisenberg model and it was shown that the 
CCM results are in good
agreement with the 
DMRG data \cite{bursill}. Hence, in the present paper 
we use the
CCM following the lines of Ref.~\onlinecite{bursill} but extend the CCM
calculations  by including the inter-chain coupling $J_\perp$.
  
The frustrated spatially anisotropic 2D 
$J_1$-$J_2$-$J_{\perp}$ 
spin-1/2-Heisenberg model
considered here reads 
\bea\label{eq1.1}H&=& \sum_n \Big \{ 
\sum_i \big [
J_1{\bf s}_{i,n}\cdot{\bf s}_{i+1,n} + J_2 {\bf s}_{i,n}\cdot{\bf
s}_{i+2,n}\big] \Big \}\nonumber\\
&+& \sum_i\sum_n J_{\perp} {\bf s}_{i,n}\cdot{\bf s}_{i,n+1} ,
\eea
where the index $n$ labels the chains and  $i$ the lattice sites within 
a chain $n$. The
NN in-chain coupling $J_1$ is fixed to either $J_1= 1$ (antiferromagnetic)
or  $J_1= -1$ (ferromagnetic). The 
inter-chain coupling $J_{\perp}$ and the frustrating NNN in-chain coupling $J_2$  are
considered as 
varying  parameters of the model.
Note that in the case considered here 
the inter-chain coupling does not lead to frustration 
and it is practically of arbitrary strength, 
i.e.\ including also the region 
beyond the 
quasi-1D limit
$|J_{\perp}| \ll |J_1 | , J_2$.
Anyhow, the opposite limit 
$|J_{\perp}| \gg |J_1| , J_2$
will not be considered for reasons of lacking physical
relevance (to the best of our knowledge). 
We focus on the extreme quantum case, i.e.\ $({\bf s}_{i,n})^2 =s(s+1)$ with
$s=1/2$.
The main point  which 
will be considered here is
the influence of the  
inter-chain coupling $J_\perp$ 
on the transition point between the collinear phase and the non-collinear
spiral phase and on the pitch angle characterizing the spiral correlations
in the quantum model with $s=1/2$.

\section{The coupled Cluster Method (CCM)}
\label{ccm}

In this section we outline only some main features of the CCM which are 
relevant for the 
model under consideration.
Again we mention that we follow the lines described in 
Ref.~\onlinecite{bursill}
where the CCM was applied to the strictly 1D problem.
For more details of the method the interested reader is referred to
Refs.~\onlinecite{bursill,bishop91,bishop91a,zeng98,bishop00,krueger00,krueger01,
ivanov02,farnell04,rachid05,rachid06,Bi:2008_j1j2j3_spinHalf,Bi:2008_j1j2j3_spinOne}.
Special attention to the CCM treatment of non-collinear GS's was paid in
Refs.~\onlinecite{bursill,krueger00,krueger01,ivanov02,rachid05,rachid06}.

The starting point for the CCM calculation is the choice of a normalized 
reference or model state
$|\Phi\rangle$, together with a complete set of (mutually commuting) 
multi-configurational creation 
operators $\{ C_L^+ \}$ and 
the corresponding set of their Hermitian adjoints $\{ C_L \}$,
\begin{eqnarray}
\label{eq2.1} \langle \Phi|C_L^+ = 0 = C_L|\Phi \rangle \quad \forall L\neq 0, \quad C_0^+\equiv 1 \\
\label{eq2.2}[C_L^+,C_{K}^+] = 0
=[C_L,C_{K}] \; .
\end{eqnarray}
With the set $\{|\Phi\rangle, C_L^+\}$ the CCM parametrization of the ket GS eigenvector
$|\Psi\ra$  of the considered many-body system is then given  by
\begin{eqnarray}\label{eq5} 
|\Psi\ra=e^S|\Phi\ra \; , \mbox{ } S=\sum_{L\neq 0}a_LC_L^+ \; .
\end{eqnarray}
The CCM correlation operator $S$ contains the correlation coefficients $a_L$
which can be determined by the so-called set of ket equations 
\begin{eqnarray}
\label{eq6}
\langle\Phi|C_L e^{-S}He^S|\Phi\rangle = 0 \;\; ; \; \forall L\neq 0.
\end{eqnarray}
For the considered frustrated spin system we choose a reference state 
corresponding to the classical  state of the model, i.e. the ferromagnetic
state $|\downarrow\downarrow\downarrow\downarrow \cdots\rangle$ along a chain for $J_1=-1$ and small $J_2$
and 
the \Neel state $|\downarrow\uparrow\downarrow\uparrow \cdots\rangle$
for $J_1=1$ and small $J_2$, whereas for larger frustration $J_2$ 
we have to choose a non-collinear reference state with spiral
correlations along the chains  
 (i.e., pictorially, $|\Phi\rangle
=|\uparrow\nearrow\rightarrow\searrow\downarrow\swarrow\cdots\rangle$)
characterized by a pitch angle $\alpha$,  i.e.\  
$| \Phi\rangle = |\Phi(\alpha )\rangle $.
Such states include the ferromagnetic state ($\alpha=0$) as well as 
the \Neel state
($\alpha=\pi$).
In the quantum model the pitch angle may be different from the corresponding
classical value $\alpha_{\rm cl}$. 
Hence, we do not choose the
classical  result for the pitch angle but, rather, we consider $\alpha$ as a free
parameter in the CCM
calculation, which has to be determined by minimization of the GS energy
given in the CCM formalism by 
$E(\alpha )= \langle \Phi (\alpha )| e^{-S}H e^S | \Phi(\alpha ) \rangle $, i.e.\ 
from
$dE/d\alpha|_{\alpha=\alpha_{\rm qu}}=0$, see also Appendix, Eqs.\ (A1)-(A6).

In order to find an appropriate set of creation operators it is convenient
to perform
a rotation of the local axes of each of the spins, such that all spins in
the reference state
align in the negative $z$-direction.
This rotation by an appropriate local angle $\delta_{i,n}$ of the 
spin on lattice site
$({i,n})$
is equivalent to the spin-operator transformation
\begin{equation}
\label{eq3} \left.\begin{array}{l} s_{i,n}^x = \cos\delta_{i,n} 
{\hat s}_{i,n}^x+\sin\delta_{i,n} {\hat s}_{i,n}^z; 
\quad s_{i,n}^y = {\hat s}_{i,n}^y  \\ s_{i,n}^z = 
-\sin\delta_{i,n} {\hat s}_{i,n}^x+\cos\delta_{i,n} {\hat s}_{i,n}^z 
\end{array} \right \} .
\end{equation}
The local rotation angle $\delta_{i,n}$ can be easily 
expressed by 
the pitch angle $\alpha$ of the spiral reference state, where the relation 
$\delta_{i,n}(\alpha)$ depends on the signs of $J_1,J_\perp$ and 
the lattice 
vector ${\bf R}_{i,n}$.  

In this new set of local spin coordinates 
the reference state and the corresponding creation operators $C_L^+$ are 
given by
\begin{equation}
\label{set1} |{\hat \Phi}\ra = |\downarrow\downarrow\downarrow\downarrow\cdots\rangle \; ; \mbox{ } C_L^+ 
= {\hat s}_{i,n}^+ \, , \, {\hat s}_{i,n}^+{\hat s}_{j,m}^+ \, , \, 
{\hat s}_{i,n}^+{\hat s}_{j,m}^+{\hat s}_{k,l}^+ \, ,\, \ldots \; ,
\end{equation}
where the indices $({i,n}),({j,m})({k,l}),\ldots$ denote arbitrary lattice sites.
In the new rotated coordinate frame the Hamiltonian (\ref{eq1.1})
becomes dependent on the pitch angle $\alpha$. It reads
\begin{widetext}
\begin{eqnarray}\label{hamiltonian_trafo}
\nonumber H&=&  \frac{J_1}{4}\sum_{i,n} [\cos(\alpha)+1]({\hat s}_{i,n}^+{\hat s}_{i+1,n}^-+{\hat s}_{i,n}^-{\hat s}_{i+1,n}^+)+
[\cos(\alpha)-1]({\hat s}_{i,n}^+{\hat s}_{i+1,n}^++{\hat s}_{i,n}^-{\hat s}_{i+1,n}^-) + 2\sin(\alpha)[{\hat s}_{i,n}^+{\hat s}_{i+1,n}^z \\ \nonumber &&\mbox{ }-{\hat s}_{i,n}^z {\hat s}_{i+1,n}^+  +{\hat s}_{i,n}^-{\hat s}_{i+1,n}^z-{\hat s}_{i,n}^z{\hat s}_{i+1,n}^-]+4\cos(\alpha){\hat s}_{i,n}^z{\hat s}_{i+1,n}^z\\ \nonumber 
&+&\frac{J_2}{4}\sum_{i,n} [\cos(2\alpha)+1]({\hat s}_{i,n}^+{\hat s}_{i+2,n}^-+{\hat s}_{i,n}^-{\hat s}_{i+2,n}^+)+ 
[\cos(2\alpha)-1]({\hat s}_{i,n}^+{\hat s}_{i+2,n}^++{\hat s}_{i,n}^-{\hat s}_{i+2,n}^-) + 2\sin(2\alpha)[{\hat s}_{i,n}^+{\hat s}_{i+2,n}^z
\\ \nonumber && \mbox{ }-{\hat s}_{i,n}^z {\hat s}_{i+2,n}^++{\hat s}_{i,n}^-{\hat s}_{i+2,n}^z-{\hat s}_{i,n}^z{\hat s}_{i+2,n}^-]+  4\cos(2\alpha){\hat s}_{i,n}^z{\hat s}_{i+2,n}^z \\ 
&-&\frac{J^A_{\perp}}{2}\sum_{i,n} ({\hat s}_{i,n}^+{\hat s}_{i,n+1}^+ +{\hat s}_{i,n}^-{\hat s}_{i,n+1}^-+ 2{\hat s}_{i,n}^z{\hat s}_{i,n+1}^z)
+  \frac{J^F_{\perp}}{2}\sum_{i,n} ({\hat s}_{i,n}^+{\hat s}_{i,n+1}^- +{\hat s}_{i,n}^-{\hat s}_{i,n+1}^+
+ 2{\hat s}_{i,n}^z{\hat s}_{i,n+1}^z),
\end{eqnarray}
\end{widetext}
where ${\hat s}_{i,n}^{\pm}\equiv {\hat s}_{i,n}^x\pm \im {\hat s}_{i,n}^y$, 
and 
$\alpha$ ($\equiv\delta_{{i,n}}-\delta_{i+1,n}$) 
is the pitch angle between the two
neighboring spins in a chain interacting via the NN bond $J_1$, which has to
be determined for the quantum model.   
For ferromagnetically coupled spin chains 
($J_{\perp}^F\not =0$) one has to set $J_{\perp}^A =0$ in
Eq.~(\ref{hamiltonian_trafo}) and vice versa. 
Therefore, from Eq.~(\ref{hamiltonian_trafo}) it is obvious that
in the quantum case considered here the role of the (unfrustrated) inter-chain coupling
$J_\perp$
introduced in
Eq.~(\ref{eq1.1}) 
is different for ferromagnetic and antiferromagnetic
$J_\perp$.
By
contrast, in the classical case
a corresponding inter-chain coupling does not 
affect the pitch angle at all.

\begin{figure}
\scalebox{0.7}{\includegraphics{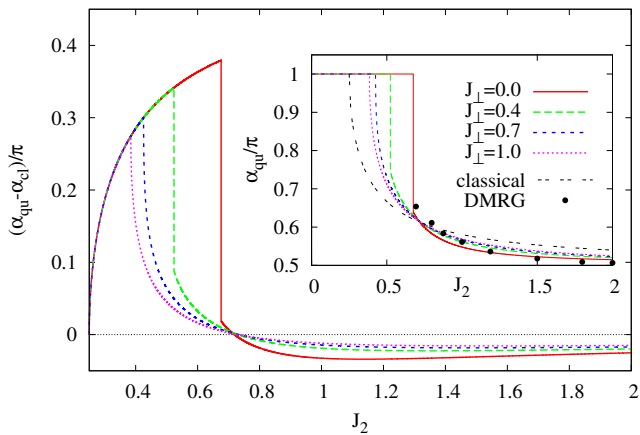}}
\caption{\label{fig1} The quantum pitch angle $\alpha_{\rm qu}$ versus 
$J_2$  for  
antiferromagnetic $J_1$ and $J_{\perp}$ 
(case (i)). 
The main panel shows the difference between  
the quantum pitch angle
$\alpha_{\rm qu}$ and its 
classical counterpart  
$\alpha_{\rm cl}$ for various inter-chain couplings $J_\perp$.
The inset shows the corresponding data for the quantum pitch angle 
$\alpha_{\rm qu}$ itself. 
For comparison the   
classical pitch angle and the DMRG data of White and Affleck~\cite{white96} for 
the strictly 
1D quantum model are also shown. 
}
\end{figure}
\begin{figure}
\scalebox{0.7}{\includegraphics{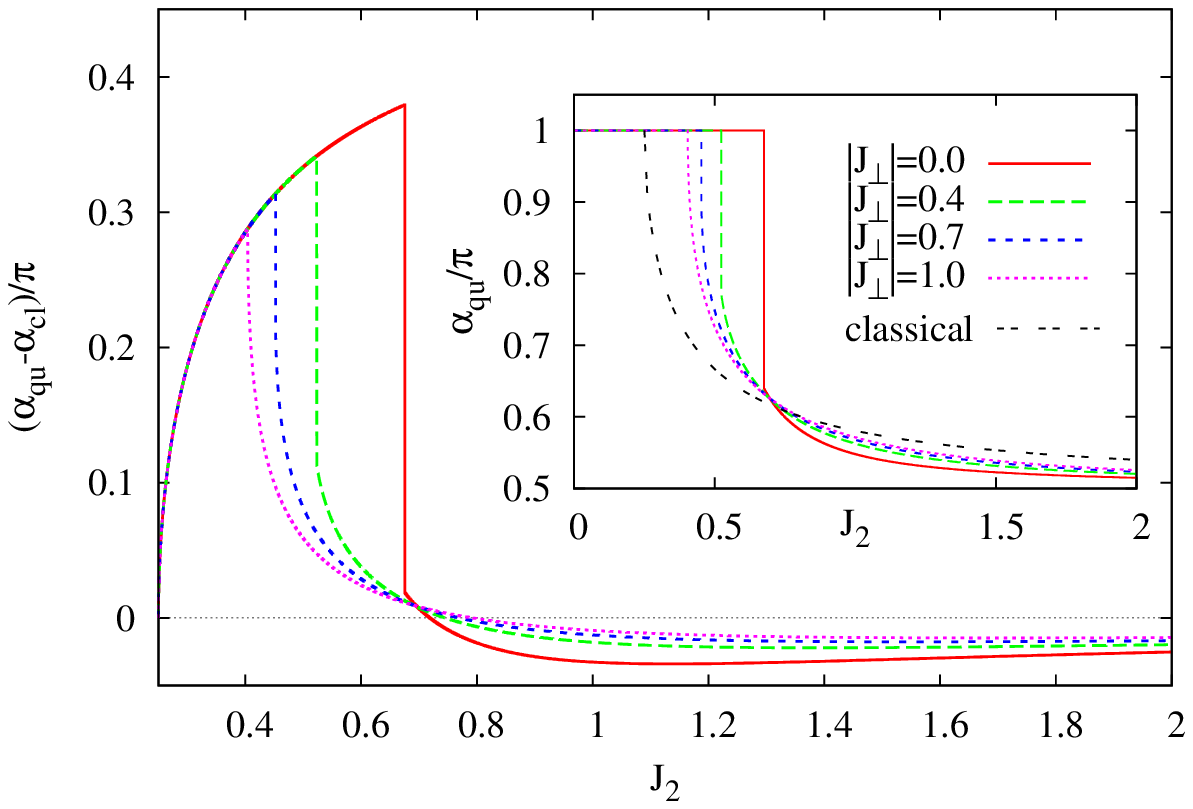}}
\caption{\label{fig2} The quantum pitch angle $\alpha_{\rm qu}$ versus $J_2$  
for  
antiferromagnetic $J_1$ and ferromagnetic $J_{\perp}$ 
(case (ii)). The main panel shows the difference between  
the quantum pitch angle
$\alpha_{\rm qu}$ and its 
classical counterpart  
$\alpha_{\rm cl}$ for various inter-chain couplings $J_\perp$.
The inset shows the corresponding data for the quantum pitch angle 
$\alpha_{\rm qu}$ itself. 
For comparison the   
classical pitch angle 
is also shown. 
}
\end{figure}
\begin{figure}
\scalebox{0.63}{\includegraphics{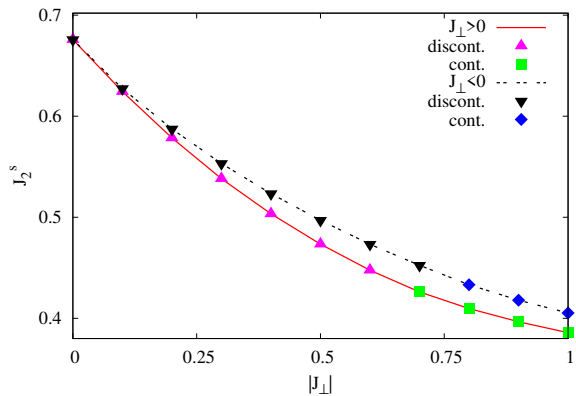}}
\caption{\label{fig3} 
The transition point $J_2^s$ as a function 
of the inter-chain coupling $J_\perp$ for $J_1=1$ and $J_\perp > 0$ (case (i))
as well as  $J_\perp
< 0$ (case (ii)).  
The triangles indicate a
discontinuous and the sqaures a continuous  
change  of
$\alpha_{\rm qu}$.}
\end{figure}

The CCM formalism would  be exact if we could take into account all possible 
multispin configurations in  the correlation operator $S$  which  is  impossible to do in practice 
for  a quantum many-body system. Hence, it is necessary to truncate the expansions of
$S$. 
In \cite{bursill} it was demonstrated, that the so-called SUB2-3 approximation for
the strictly 1D system 
leads to results of comparable accuracy to those obtained using the DMRG method. 
In the SUB2-3 approximation all configurations are included which span a 
range of no more than 
$3$ contiguous sites and contain only  $2$ or fewer spins. 
A particular advantage of the SUB2-3 approximation consists of the
possibility to find the relevant CCM equations (\ref{eq6}) in closed analytical
form, see
Appendix \ref{appendix}. These explicit  equations provided here
can be used to find the
quantum pitch angle $\alpha_{\rm qu}$ for an
arbitrary set of parameters $J_1$, $J_2$, $J_\perp$ by simple numerical
solution of them.

\section{Results}

\begin{figure}
\scalebox{0.63}{\includegraphics{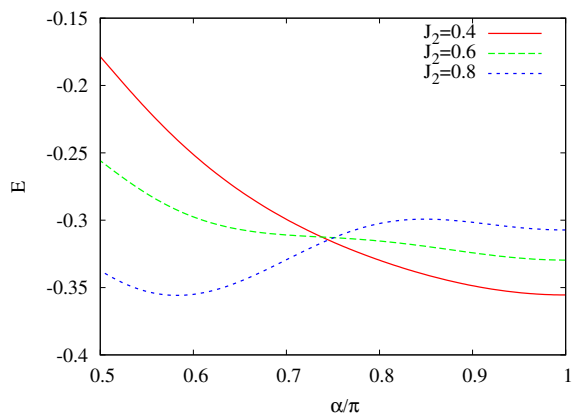} }
\caption{\label{fig4}The GS energy versus  pitch angle $\alpha$ for different (fixed) $J_2$  
and antiferromagnetic $J_1$  ($J_{\perp}=0$).}
\end{figure}

\begin{figure}
\scalebox{0.7}{\includegraphics{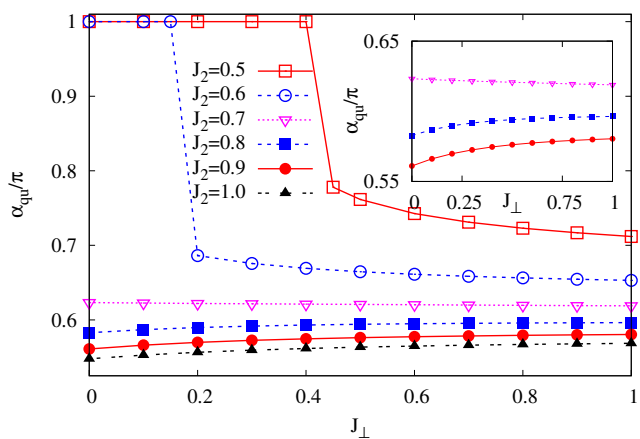} }
\caption{\label{fig5} The quantum pitch angle $\alpha_{\rm qu}$ versus
$J_\perp$
for  antiferromagnetic $J_1=1$ (case (i)) and various values of the frustrating
NNN exchange $J_2$ (the lines are guides for the eyes). 
The inset shows the data for $J_2=0.7, 0.8, 0.9$ with an enlarged scale.}    
\end{figure}

For the model under consideration we have calculated 
that  point
$J_2^{s}$ where the GS state spin-spin correlations change from collinear to non-collinear 
spiral correlations as well as the 
quantum pitch angle $\alpha_{\rm qu}$. 
In what follows we call the point $J_2^{s}$ the 'transition point'.  We
mention, however, 
that the question for magnetic GS long-range order goes
beyond the scope of the present paper. Generally one can argue that for the
strictly 1D problem the GS (except the simple ferromagnetic state) does 
not exhibit
magnetic long-range order, whereas for finite $J_\perp$ GS long-range order
can exist, cf. e.g. Refs.~\onlinecite{affleck,wang,sandvik,kim,zinke08}.
In particular,
the GS phase for
larger $J_2$ and small $J_\perp$  is magnetically disordered and 
may have a weak
spontaneous dimerization
along with finite-range incommensurate magnetic correlations \cite{affleck}. 
Hence the 'transition point' $J_2^{s}$ for small $J_\perp$ 
may locate that narrow para\-meter region where finite-range magnetic
correlations are changing from commensurate to incommensurate ones, 
but does not indicate a true
quantum phase transition.

We present data for $J_\perp = \pm 0, 0.1, 0.2, \ldots, 1.0$ and a fine
net of $J_2$ values. 
 For the sake of clarity in the following we asort  the  results into four
cases, depending on the signs of $J_1$ and $J_{\perp}$, namely 
(i) $J_1 =1$, $J_\perp \ge 0$; 
(ii) $J_1 =1$, $J_\perp \le 0$; 
(iii) $J_1 =-1$, $J_\perp \ge 0$; 
(iv) $J_1 =-1$, $J_\perp \le 0$.

Since for cases (i) and (ii) the behavior is quite similar, we can discuss both
cases together. The pitch angle in dependence 
on $J_2$ 
is shown in Figs.~\ref{fig1} (case (i))
and \ref{fig2} (case (ii)). 
For
comparison we also draw the corresponding DMRG data of White and Affleck~\cite{white96} for
the strictly 1D problem in Fig.~\ref{fig1}.
These data agree quite well with the CCM data in particular for larger
$J_2$.
It is obvious that quantum fluctuations change the classical
correlations drastically. 
In particular, by contrast  to the classical case,   
the collinear quantum state can survive into the region 
$J_2 > J_2^{s, \rm cl}=0.25$,
where classically it is already unstable.
This effect is known as {\it order from disorder} \cite{villain,shender} and
is widely observed in quantum spin systems \cite{henley89,Kubo_JPSJ,krueger00,rachid05}.   
We find, e.g.  $J_2^{s} \approx 0.68 $
for the quantum model with $J_\perp=0$, which is in
good agreement with known results, see, e.g.,  Fig.~3 in Ref.~\onlinecite{aligia00}.

\begin{figure}
\scalebox{0.7}{\includegraphics{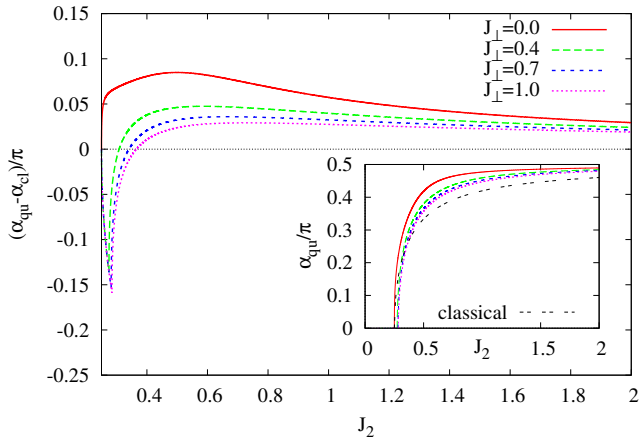} }
\caption{\label{fig6} The quantum pitch angle $\alpha_{\rm qu}$ versus $J_2$  for  
ferromagnetic $J_1$ and antiferromagnetic $J_{\perp}$ 
(case (iii)). 
The main panel shows the difference between  
the quantum pitch angle
$\alpha_{\rm qu}$ and its 
classical counterpart  
$\alpha_{\rm cl}$ for various inter-chain couplings $J_\perp$.
The inset shows the corresponding data for the quantum pitch angle 
$\alpha_{\rm qu}$ itself. 
For comparison the  classical pitch angle is also shown. 
}
\end{figure}

\begin{figure}
\scalebox{0.7}{\includegraphics{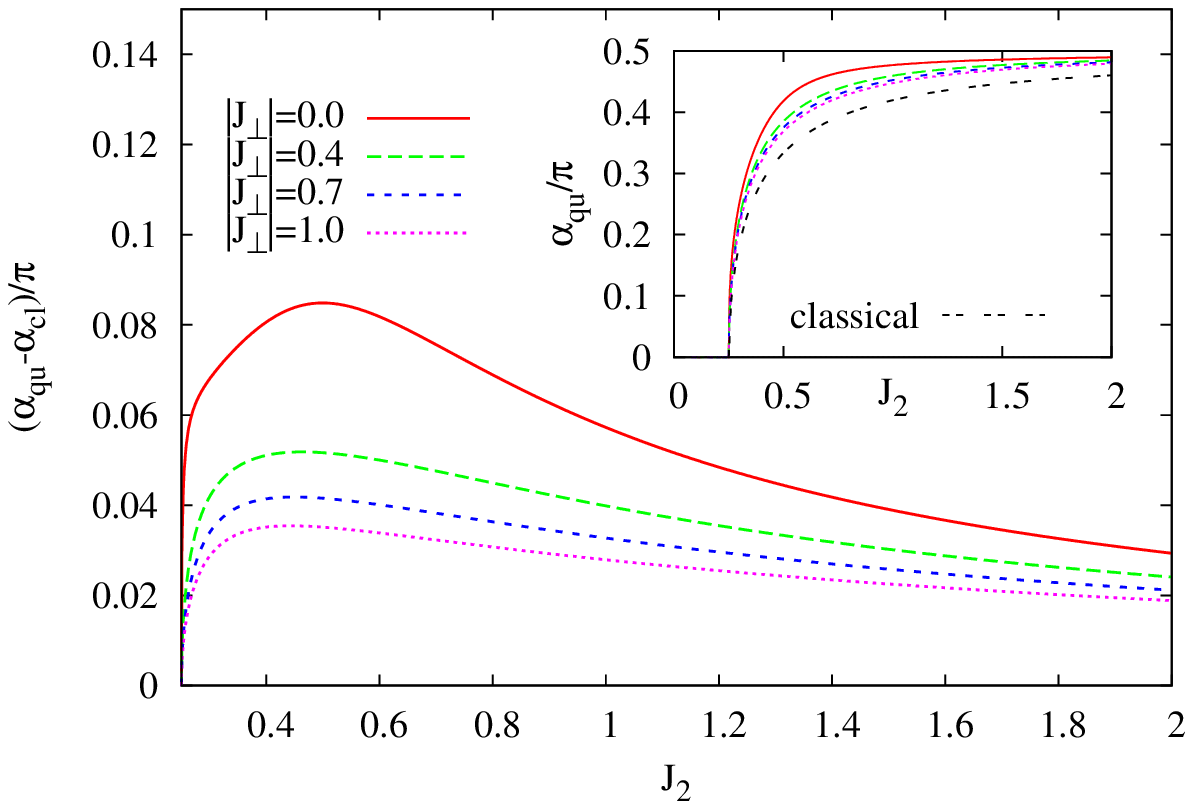} }
\caption{\label{fig7} The quantum pitch angle $\alpha_{\rm qu}$ 
versus $J_2$  for  
ferromagnetic $J_1$ and $J_{\perp}$ 
(case (iv)). 
The main panel shows the difference between  
the quantum pitch angle
$\alpha_{\rm qu}$ and its 
classical counterpart  
$\alpha_{\rm cl}$ for various inter-chain couplings $J_\perp$.
The inset shows the corresponding data for the quantum pitch angle 
$\alpha_{\rm qu}$ itself. 
For comparison the  classical pitch angle is also shown. 
}
\end{figure}

\begin{figure}
\scalebox{0.63}{\includegraphics{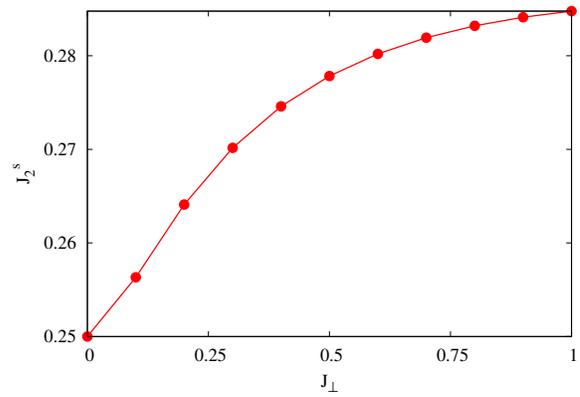} }
\caption{\label{fig8} 
The transition point $J_2^s$ as a function 
of the inter-chain coupling $J_\perp$ for $J_1=-1$ and $J_\perp > 0$ (case
(iii)).}
\end{figure}

\begin{figure}
\scalebox{0.6}{\includegraphics{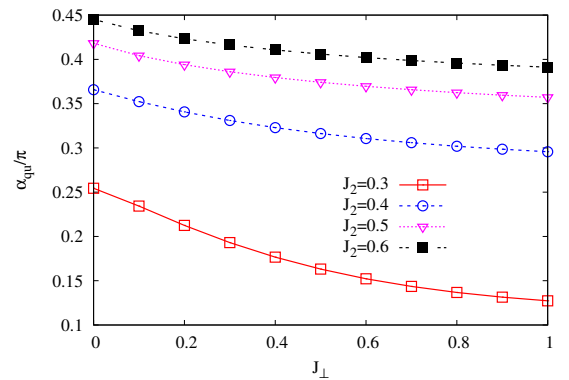} }
\caption{\label{fig9} The quantum pitch angle $\alpha_{\rm qu}$ versus
$J_\perp$
for  ferromagnetic $J_1=-1$  (case (iii)) and various values of the frustrating
NNN exchange $J_2$. }
\end{figure}

Switching on $J_\perp$, i.e. increasing the dimension of the spin system, 
the effect of quantum fluctuations should become weaker and, as a result,
the
collinear quantum state gives way for the state with incommensurate correlations
at smaller values of $J_2$, i.e. the 
transition point moves towards the classical value, see 
Fig.~\ref{fig3}. 
However, even in nearly 
isotropic 2D systems ($J_\perp \sim 1$)
the quantum fluctuations are still 
important and one has  $J_2^s \approx 0.38$ ($J_2^s \approx 0.40$) for case (i)
(for case (ii)) which is 
significantly above the classical value.

The shift of the transition point leads to an interesting behaviour of the
difference $\alpha_{\rm qu}
-\alpha_{\rm cl}$. 
It is positive for  $0.25 < J_2 < J_2^*$ but negative
for $J_2 > J_2^*$.
Surprisingly we find $J_2^*\approx 0.7 $  
as almost independent of $J_\perp$, i.e. all curves 
$\alpha_{\rm qu} -\alpha_{\rm cl}$ vs. $J_2$ meet approximately in one point, see
Figs.~\ref{fig1} and \ref{fig2}.
Note that for large $J_2 > J_1$ 
the model with strongest
quantum fluctuations, i.e. $J_\perp=0$, approaches the
limit $\alpha_{\rm qu} \to \pi/2$ most rapidly.

Next we discuss the 
behavior
of the pitch angle at the transition point.
For smaller values of the inter-chain coupling ($J_\perp \lesssim 0.7$ for case (i)
and $J_\perp \lesssim 0.8 $ for case (ii)) we find a
discontinuous 
behavior 
of the quantum pitch angle. A similar jumpwise change of
the pitch angle has been found  for 2D frustrated quantum spin
models~\cite{krueger00,rachid05}. Note that in Ref.~\onlinecite{bursill} 
the CCM curve was terminated before the jump occurs so that the
jump  has not been observed
there~\cite{damian}.  
The discontinuous change of $\alpha_{\rm qu}$ is related to  
 the existence of two
minima in the $E$ versus $\alpha$ curve and takes place for parameter values
where both minima have equal depth, cf. Fig.~\ref{fig4}. 
For larger 
$J_\perp$ (weaker quantum fluctuations) $E(\alpha)$ exhibits only one minimum 
and the quantum pitch angle is changing continuously 
from the
collinear to the spiral GS. 
We remind the reader
that for the classical case 
the transition from the commensurate to the incommensurate GS takes place at $J_2=|J_1|/4$ and that it is continuous 
for any value of $J_\perp$.

Finally, in Fig.~\ref{fig5} 
we have drawn the pitch angle $\alpha_{\rm qu}$
in dependence on the inter-chain coupling $J_\perp$ 
for case (i).
As discussed above, with increasing $J_\perp$ the quantum pitch angle moves
towards the corresponding classical value. 
However, in accordance with the above discussion of 
the change of the sign of $\alpha_{\rm qu} -\alpha_{\rm cl}$ 
we find two different
regimes: For $J_2 \lesssim J_2^*$ the $\alpha_{\rm qu}$ increases with growing
$J_\perp$ while for  $J_2 \gtrsim J_2^*$ the $\alpha_{\rm qu}$ decreases with 
$J_\perp$.

Let us now pass to cases (iii) and (iv), i.e. when 
$J_1=-1$ is ferromagnetic.
By contrast to cases (i) and (ii), we find 
that here 
the transition from the collinear state to the spiral state
is always continuous. 
The reason for that can be again attributed
to the strength of quantum fluctuations. 
For ferromagnetic $J_1$ the relevant
collinear state at $J_2 < J_2^s$ consists of  ferromagnetic
chains, having a classical (i.e. 'non-fluctuating') GS, coupled by $J_\perp$.
While for case (iv) the ferromagnetic inter-chain
coupling does not change this 'non-fluctuating' GS at all, for case (iii) due to
the presence of antiferromagnetic couplings quantum fluctuations become
relevant and the GS becomes a true quantum
state, but the 
change of magnetic correlations within the chains for $J_\perp > 0$ remains weak.
Hence, virtually no (case (iv)) or only weak (case (iii)) quantum fluctuations
occur at the transition from the collinear to the noncollinear GS.
As a
result the transition takes place precisely at $J_2=|J_1|/4$ for case
(iv) while for case
(iii) 
the transition point $J_2^s$ is above the classical value indicating again
an {\it order from disorder} effect. However, the shift of $J_2^s$ 
is small because the
quantum fluctuations are weak, see Figs.~\ref{fig6} and \ref{fig8}.  
Nevertheless, such a small shift of $J_2^s$ due to a finite
inter-chain coupling might be important for systems such as Li$_2$ZrCuO$_4$
and Li$_2$CuO$_2$~\cite{drechsQneu,malek}
being near the transition point.
Note that Bader and Schilling~\cite{bader} first found that the transition
point is fixed at $J_2=|J_1|/4$, 
 if the collinear state is the classical
ferromagnetic one. Note further that a similar change from a discontinuous to
a continuous transition was discussed in Ref.~\onlinecite{krueger01}.

The quantum pitch angle for cases (iii) and (iv) is shown in Figs.~\ref{fig6} and
\ref{fig7}. 
Obviously, there is also a signi\-ficant difference between the quantum and the classical pitch 
angle, however, it is smaller than for cases (i) and (ii). For $J_\perp=0$
the largest difference
of about $0.09\pi$ is
found at $J_2 \approx 0.5$. 
For case (iv) for all values of $J_\perp \le 0$ and $J_2>0.25$ 
the qantum pitch angle is larger than
the classical one. 
On the other hand, for case (iii) the shift of the transition point $J_2^s$ leads to a change  
of the sign of 
$\alpha_{\rm qu}-\alpha_{\rm cl}$, i.e. the antiferromagnetic inter-chain
coupling yields are more
subtil change of the GS correlations by quantum fluctuations.  
Moreover, due to the shift of $J_2^s$  we find a quite large  difference
$|\alpha_{\rm qu}-\alpha_{\rm cl}|$ for $J_\perp > 0$  and $J_2\approx
0.27 \ldots 0.28$. 

Similarly as for cases (i) and (ii) 
we observe that for large $J_2 > J_1$ 
the model with strongest
quantum fluctuations, i.e. $J_\perp=0$, approaches the
limit $\alpha_{\rm qu} \to \pi/2$ most rapidly.
We mention here that a large value of $|J_2/J_1|$ is realized 
e.g. in LiCuVO$_4$ \cite{gibson,ender,drechs4}, for which $J_2/J_1 \approx -2.4$ 
has been estimated \cite{ender,drechs4}.
The variation of $\alpha_{\rm qu}$ with $J_\perp$ for various $J_2$ is shown
in Fig.~\ref{fig9}. It can be seen that the variation of $\alpha_{\rm qu}$
with $J_\perp$ is largest for small $J_\perp$.
Furthermore, in difference to cases (i) and (ii) $\alpha_{\rm qu}$
decreases monotonously
with $J_\perp$ for all values of $J_2> J_2^s$.

\begin{table}[h]
\caption{\label{table1} Numerical values for the  exponent $\nu$ and the
transition point 
$J_2^s$ in dependence on $J_{\perp}$ for the cases (i)-(iv), cf.
Eq.~(\ref{a_qu}).  }\vspace{0.5cm}
\begin{tabular}{|c|c|c|c|c|}\hline
$\mbox{ }|J_{\perp}|$\mbox{ } & \multicolumn{2}{|c} { (i): $J_1=1, J_{\perp}>0$} & \multicolumn{2}{|c|}{ (ii): $J_1=1, J_{\perp}<0$}\\ \hline
& $J_2^s$ & $\nu$ & $J_2^s$ & $\nu$\\ \hline
0.70  &  0.426  &  1.364  &  -  &  -  \\ \hline
0.80  &  0.409  &  1.317  &  0.433  &  1.341  \\ \hline
0.90  &  0.397  &  1.283  &  0.418  &  1.294  \\ \hline
1.00  &  0.386  &  1.256  &  0.405  &  1.258  \\ \hline
& \multicolumn{2}{|c} {(iii): $J_1=-1, J_{\perp}>0$} & \multicolumn{2}{|c|}{(iv): $J_1=-1, J_{\perp}<0$}\\ \hline
& $J_2^s$ & $\nu$ & $J_2^s$ & $\nu$\\ \hline
0.00  &  0.250  &  1.972  &  0.250  &  1.972  \\ \hline
0.10  &  0.256  &  1.761  &  0.250  &  1.742  \\ \hline
0.20  &  0.264  &  1.662  &  0.250  &  1.630  \\ \hline
0.30  &  0.270  &  1.588  &  0.250  &  1.552  \\ \hline
0.40  &  0.275  &  1.530  &  0.250  &  1.494  \\ \hline
0.50  &  0.278  &  1.483  &  0.250  &  1.448  \\ \hline
0.60  &  0.280  &  1.443  &  0.250  &  1.411  \\ \hline
0.70  &  0.282  &  1.410  &  0.250  &  1.380  \\ \hline
0.80  &  0.283  &  1.381  &  0.250  &  1.354  \\ \hline
0.90  &  0.284  &  1.357  &  0.250  &  1.331  \\ \hline
1.00  &  0.285  &  1.336  &  0.250  &  1.312  \\ \hline
\end{tabular}
\end{table}

\begin{figure}
\scalebox{0.6}{\includegraphics{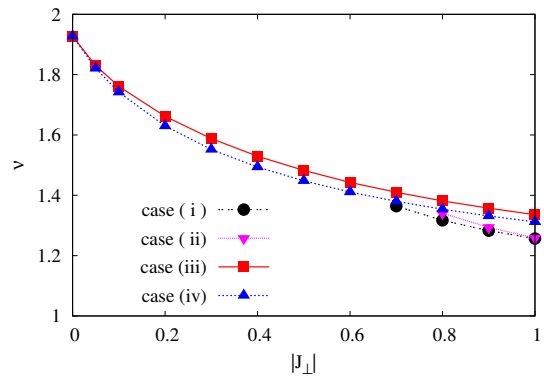} }
\caption{\label{fig10} The  exponent $\nu$, see
Eq.~(\ref{a_qu}), in dependence on $J_{\perp}$ for 
the cases (i)-(iv). }
\end{figure}

As mentioned in Sect.~\ref{intro}, a 
classical relation like Eq.~(\ref{a_cl})
 between the exchange couplings
and the   pitch angle 
has been used to discuss the $J_2/J_1$ ratio in 
Refs.~\onlinecite{capogna2005,mazurenko2007}
this way ignoring quantum effects.
To overcome this problem we will provide 
an empirical formula that fits the continuous part of the
$\alpha_{\rm qu}(J_2)$ given in Figs.~\ref{fig1}, \ref{fig2}, \ref{fig6},
\ref{fig7} very well.
Having in mind that the shape of the continuous part of the
$\alpha_{\rm qu}(J_2)$ resembles the classical behavior        
we find that $\alpha_{\rm qu}(J_2)$
is well approximated by (written now in dimensional
exchange units for the convenience of application in real experimental 
situations): 
\begin{equation}\label{a_qu}
\alpha_{\rm qu}(J_2) =
\arccos\left(\frac{ - J_1}{[4(J_2-J_2^s+\frac{1}{4}|J_1|)]^{\nu}}\right)
\end{equation}
with the exponent $\nu$ as fitting parameter. Obviously, 
Eqs.~(\ref{a_qu}) and
(\ref{a_cl}) coincide for $J_2^s=|J_1|/4$ 
and $\nu=1$.   
 In
Fig.~\ref{fig10} we show $\nu$ in dependence on $J_{\perp}$ for the four cases (i)-(iv). 
We find that $\nu$ is always larger than the classical value $\nu_{\rm cl}=1$.
In accordance with the above discussion,  $\nu$ decreases  
with increasing $J_{\perp}$, i.e. it goes towards the classical exponent
$\nu_{\rm cl}$. 
From the experimental point of view the edge-shared chain cuprates are of
particular interest. The parameter situation of these compounds corresponds
to case (iii). 
Hence, we give
here simple fit formulas for that case which describe the behavior of $J_2^s(J_\perp)$ as shown in
Fig.~\ref{fig8} and the behavior of $\nu(J_\perp)$ as shown in
Fig.~\ref{fig9} by the red line.
We find that 
\begin{equation} \label{fit}
\nu =\frac{a}{(\frac{J_\perp}{|J_1|}+b)^c}\;  , 
\; J_2^s= \frac{|J_1|}{4} + p \hskip 1pt
|J_1|\hskip 1pt
\tanh \left(q \frac{J_\perp}{| J_1 |} \right),
\end{equation}
with $a=1.37$, 
$b=0.13$, $c=0.17$, $p=0.036$, and $q=2.11$ 
provides  a reasonable fit of our data for case (iii).
In addition, 
we present numbers for $J_2^s$ and $\nu$ for various values of 
$J_\perp$ for all cases in
table~\ref{table1}. The fitting formulas (\ref{fit}) as well as the data in
table~\ref{table1} can be used in combination with Eq.~(\ref{a_qu}) 
to fix the $J_2/J_1$ ratio using the pitch angle 
$\alpha$ as an input, e.g. from neutron
scattering \cite{gibson,matsuda,ender,capogna2005}.

Finally, we note that recently the pitch angle of 
Li$_2$ZrCuO$_4$ has been determined from $^7$Li-NMR-data to amount 
$\alpha = 33^\circ \pm 2^\circ $ \cite{tarui}. This value corresponds 
to a predicted ratio $-J_2/J_1= 0.298$ within the framework of the
classical Eq.\ (1). This value is surprisingly very close to the 
ratio $-J_2/J_1$= 0.3 estimated 
from thermodynamic properties within the 1D-{\it quantum} spin-1/2 
$J_1$-$J_2$-model 
\cite{drechsQneu}. However, from Fig.~\ref{fig9} for a realistic
weak effective antiferromagnetic 
inter-chain coupling $J_{\perp}\lesssim 0.1 |J_1|$ 
a somewhat {\it larger} pitch angle would be predicted, namely 
$\alpha \approx 42^\circ $ for $-J_2/J_1$= 0.3. 
Hence, other factors such as a sizable
exchange anisotropy are expected to be relevant in
this material. Such an anisotropy may lead to a modification of the
classical spiral as well as to a reduction of 
quantum fluctuation compared with the
isotropic spin model considered here.
The 
corresponding effects are outside 
the scope of the present paper. They will be considered 
elsewhere.

\section{Summary}
Based on the  coupled cluster method (CCM) we have studied the GS correlations of
a 2D array of frustrated spin-$\frac{1}{2}$ $J_1$-$J_2$ chains coupled by an
inter-chain exchange interaction $J_\perp$.
We have discussed 
the influence of quantum fluctuations, frustration  and inter-chain 
coupling on the pitch angle
and the transition point between a GS with collinear commensurate
correlations and a GS with incommensurate spiral correlations. 
Using the  CCM within the so-called SUB2-3 approximation we obtain a closed
set of analytical equations which can be used to calculate the pitch angle
for an arbitrary set of exchange parameters.
We have found that for $J_2 > 0.25 |J_1|$  the pitch angle of the quantum
model significantly deviates 
from the classical value and can be strongly influenced by the inter-chain 
coupling
$J_\perp$. Furthermore, we have observed that the quantum pitch angle   
approaches its limiting value $\pi/2$ for increasing values of $J_2$ 
much faster than 
for the classical model. For several combinations 
of the sign of $J_1$ and $J_{\perp}$ we have found that the
change of the pitch angle within the quantum model could be  discontinuous while  
the change of the pitch angle  within the classical model is always
continuous.\\

{{\it   Acknowledgment:}
The authors thank J.~Schulenburg and R.~Kuzian
for helpful  discussions. This work was supported by the DFG (project
RI 615/16-1, DR 269/3-1).}\\

\appendix
\section{The CCM SUB2-3 approximation}
\label{appendix}

\begin{figure}[h]
\scalebox{0.2}{\includegraphics{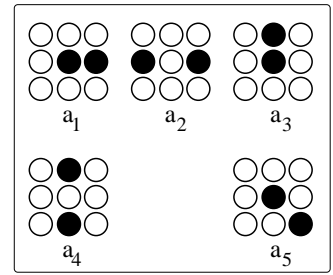} }
\caption{\label{fig11} Illustration of the five 
configurations on the spin lattice which contribute to the CCM SUB2-3
approximation. Each configuration is related to a particular 
multi-configurational creation 
operators $\{ C_L^+ \}$
and   to the corresponding  correlation coefficient
$a_L$, $L=1, \ldots, 5$, see Eqs. (\ref{eq5}) and (\ref{set1}). The circles 
in the figure represent lattice sites, the black circles 
indicate the position of the flipped spins in a certain 
configuration.}
\end{figure}
Within the SUB2-3 approximation scheme the CCM correlation operator
$S$, see Eq.~(\ref{eq5}), contains five non-equivalent  correlation coefficients
$a_L$, $L=1, \ldots, 5$,
corresponding to the lattice configurations shown in Fig.~\ref{fig11}. 
These configurations (or lattice animals) represent the arrangement of 
spin operators acting on the lattice spins. 
To each configuration belongs a corresponding ket equation, see
Eq.~(\ref{eq6}).
The set of these ket equations can be found by a bit tedious but 
straightforward     
calculation. 
It reads
{\small 
\begin{widetext}
\begin{eqnarray}
\label{sub2-3eqn1}\nonumber\noindent
&&\frac{J_1}{4}\bigl \{ [\cos(\alpha)-1](1-12a_1^2+8a_2^2+8a_3^2+8a_4^2+24a_5^2)+4a_2[\cos(a)+1]-8a_1\cos(\alpha)\bigr\}+ \frac{J_2}{4}\bigl \{ [\cos(2\alpha)-1](16a_3a_5 \\ 
&&-16a_1a_2)+4a_1[\cos(2\alpha)+1]-16a_1\cos(2\alpha)\bigr \}+ \frac{J_{\perp}^F}{2}(8a_5-8a_1)-\frac{J_{\perp}^A}{2}(16a_4a_5-16a_1a_3+16a_2a_5 -8a_1)=0\\\nonumber
\\ 
\label{sub2-3eqn2}\nonumber
&&\frac{J_1}{4}\bigl \{ [\cos(\alpha)-1](-16a_1a_2+16a_3a_5)+4a_1[\cos(a)+1]-16a_2\cos(\alpha)\bigr\}+ \frac{J_2}{4}\bigl \{ [\cos(2\alpha)-1](1-12a_2^2+8a_1^2 \\ 
&&+8a_3^2+8a_4^2+16a_5^2)-8a_2\cos(2\alpha)\bigr \}- \frac{J_{\perp}^F}{2}(8a_2)-\frac{J_{\perp}^A}{2}(16a_1a_5-16a_2a_3 -8a_2)=0\\\nonumber
\\ 
\label{sub2-3eqn3}\nonumber
&&\frac{J_1}{4}\bigl \{ [\cos(\alpha)-1](-16a_1a_3+16a_2a_5+16a_4a_5)+8a_5[\cos(a)+1]-16a_3\cos(\alpha)\bigr\}+ \frac{J_2}{4}\bigl \{ [\cos(2\alpha)-1](-16a_2a_3 \\ 
&&+16a_1a_5)-16a_3\cos(2\alpha)\bigr \}+
\frac{J_{\perp}^F}{2}(4a_4-4a_3)-\frac{J_{\perp}^A}{2}(1-12a_3^2+8a_1^2+8a_2^2+8a_4^2+24a_5^2-4a_3)=0\\\nonumber
\\ 
\label{sub2-3eqn4}\nonumber
&&\frac{J_1}{4}\bigl \{ [\cos(\alpha)-1](-16a_1a_4+16a_3a_5)-16a_4\cos(\alpha)\bigr\}+ \frac{J_2}{4}\bigl \{ [\cos(2\alpha)-1](-16a_2a_4+8a_5^2) \\ 
&&-16a_4\cos(2\alpha)\bigr \}+ \frac{J_{\perp}^F}{2}(4a_3-8a_4)-\frac{J_{\perp}^A}{2}(-16a_3a_4+16a_1a_5 -8a_4)=0\\\nonumber
\\ 
\label{sub2-3eqn5}\nonumber
&&\frac{J_1}{4}\bigl \{ [\cos(\alpha)-1](16a_2a_3-16a_1a_5+16a_3a_4)+8a_3[\cos(a)+1]-32a_5\cos(\alpha)\bigr\}+ \frac{J_2}{4}\bigl \{ [\cos(2\alpha)-1](16a_1a_3-32a_2a_5 \\ 
&&+16a_4a_5)+8a_5[\cos(2\alpha)+1]-32a_5\cos(2\alpha)\bigr \}+ 
\frac{J_{\perp}^F}{2}(8a_1-16a_5)-\frac{J_{\perp}^A}{2}(16a_1a_2-16a_3a_5+16a_1a_4 -16a_5)=0 \; .
\end{eqnarray}
\end{widetext}
}
The GS energy is a function  of (some of) 
these correlation coefficients and of the pitch angle $\alpha$. It reads 
\begin{widetext}
\begin{eqnarray}
\label{sub2-3eqn6} &&E=\frac{J_1}{4}\bigl \{ 2a_1 [\cos(\alpha)-1] + \cos(\alpha)\bigr \}+\frac{J_2}{4}
\bigl \{ 2a_2 [\cos(2\alpha)-1] + \cos(2\alpha)\bigr \}+\frac{J_{\perp}^F}{4}-\frac{J_{\perp}^A}{4}(4a_3+1) \; .
\end{eqnarray}
\end{widetext}
Note that in the above equations one has to set $J_{\perp}^A =0$ ($J_{\perp}^F
=0$) and to replace $J_{\perp}^F$
by $J_\perp$ ($J_{\perp}^A$
by $J_\perp$)  for ferromagnetic (antiferromagnetic) $J_\perp$. 
To determine the quantum pitch angle $\alpha_{\rm qu}$ as a function
of the parameters $J_1$, $J_2$ and $J_\perp$
one has to solve the equation $dE/d\alpha|_{\alpha=\alpha_{\rm qu}}=0$, cf.
Sect.~\ref{ccm},
together with the set of ket equations~(\ref{sub2-3eqn1}) - (\ref{sub2-3eqn5})
selfconsistently
by standard numerics.

Finally, we will illustrate some limiting cases contained in Eqs.~(\ref{sub2-3eqn1}) - (\ref{sub2-3eqn6}).
For $J_\perp = 0$ one has $a_3=a_4=a_5=0$ and the remaining two non-trivial
Eqs.~(\ref{sub2-3eqn1}) and (\ref{sub2-3eqn2}) then  
coincide  with the corresponding equations given in Ref.~\onlinecite{bursill}.
 In case of $J_2=0$, $J_{\perp} =0$, $a_3=a_4=a_5=0$ but $J_1\not=0$ (or
alternatively, $J_2=0$, $J_{1} =0$, $a_1=a_2=a_5=0$ but $J_\perp \not=0$) one  
finds the two ket equations for the simple unfrustrated linear chain.

\end{document}